\documentclass[two-column, nofootinbib]{revtex4-1}

\usepackage{amsmath}
\usepackage{tabularx}
\usepackage{graphicx}
\usepackage[usenames,dvipsnames,svgnames]{xcolor}
\usepackage{hyperref}

\hypersetup{dvips,dvipdfm,colorlinks=true,urlcolor=magenta,filecolor=magenta,linktoc=page,citecolor=red,linkcolor=blue,bookmarks=true}

\begin{document}

\title{The Lambert $W$ function: A newcomer in the Cosmology class?}

\author{Subhajit Saha\footnote {Electronic Address: \texttt{\color{blue} subhajit1729@gmail.com}}}
\affiliation{Department of Mathematics, \\ Panihati Mahavidyalaya, Kolkata 700110, West Bengal, India.}

\author{Kazuharu Bamba\footnote {Electronic Address: \texttt{\color{blue} bamba@sss.fukushima-u.ac.jp}}}
\affiliation{Division of Human Support System, Faculty of Symbiotic Systems Science, \\ Fukushima University, Fukushima 960-1296, Japan.}


\begin{abstract}

\begin{center}
(Dated: The $12^{\text{th}}$ September, $2019$)
\end{center}

We propose a novel equation of state (EoS) which explains the evolutionary history of a flat Friedmann-Lemaitre-Robertson-Walker (FLRW) Universe. The uniqueness of this EoS lies in the fact that it incorporates the Lambert $W$ function in a special fashion. It is explicitly demonstrated that with observationally relevant values of the unknown parameters $\vartheta _\mathrm{1}$ and $\vartheta _\mathrm{2}$, all the evolutionary phases of the Universe can be reproduced. Moreover, it also shows that the initial singularity is unavoidable and asserts that the late-time acceleration of the Universe would continue forever.\\\\
Keywords: Lambert $W$ function; FLRW universe; Equation of state; Cosmic history\\\\
PACS Numbers: 98.80.-k\\\\

\end{abstract}

\maketitle


It is now well-established that our Universe is undergoing an accelerated expansion\cite{Riess1,Perlmutter1,Schmidt1,Ade1}. This startling fact has been established by analyzing observational data obtained from Supernovae Type Ia (SNe Ia), cosmic microwave background (CMB) radiation, baryon acoustic oscillations (BAO), large scale structure (LSS) of spacetime, and weak lensing. Most cosmologists have taken either of the following two approaches to interpret the observed acceleration at the present epoch. The first approach introduces an exotic substance called dark energy (DE) on the right hand side (matter part) of the EFE, while the other one modifies the left hand side (geometric part) of the EFE. For reviews on the two approaches, the reader is referred to Refs. \cite{Copeland1,Capozziello1,Nojiri2,Capozziello2,Bamba1,Bamba2,Cai1,Nojiri1}. It is worth noting that the most widely accepted model, or concordance model in other words, in modern day Cosmology is the $\Lambda$ cold dark matter ($\Lambda$CDM) model which states that the Universe contains a tiny, yet non-zero, cosmological constant $\Lambda$ which acts as DE (the dominant component), and matter in the form of dust, which together make up almost 96\% of the energy budget of the Universe. However, $\Lambda$ suffers from serious problems, most notably, the cosmological constant problem (CCP) and the coincidence problem. So, alternative DE models have come up at different times in the literature which assume that the CCP is solved in such a way that $\Lambda$ vanishes completely. Now, there are usually two ways by which a DE model can be described --- a. fluid description in which the pressure is related to the energy density through an equation of state (EoS), $w_\mathrm{eff}$, and b. scalar field description in which the energy density and pressure of the field is determined from the given action. In this letter, we propose an EoS which deals with a special mathematical function, known as the Lambert $W$ function. We model the fluid content of a flat Friedmann-Lemaitre-Robertson-Walker (FLRW) Universe with such a novel EoS and thereby compare the evolution of the Universe with observed facts.\\

The Lambert $W$ function (also known as the omega function or product logarithm) was derived and used independently by several researchers before the mathematicians and computer scientists settled on a common notation in the mid-1990s \cite{Polya1,Corless1}. The function gained a considerable attention within the mathematical community recently [http://www.orcca.on.ca/LambertW/]. There are numerous, well-documented applications of $W$ in mathematics (such as linear delay-differential equations \cite{Asl1}), numerics \cite{Corless1}, computer science \cite{Bustos-Jimenez1}, and engineering \cite{Yi1}. It also has quite a handful of applications in physics, most notably in quantum mechanics (solutions for double-well Dirac-delta potentials \cite{Scott1,Scott2,Scott3}), quantum statistics \cite{Valluri1}, solutions to (1+1)-gravity problem \cite{Scott4} and inverse of Regge-Finkelstein coordinates \cite{Regge1} in general relativity, statistical mechanics \cite{Caillol1}, fluid dynamics \cite{Pudasaini1}, optics \cite{Steinvall1}, electrostatics, quantum chromodynamics \cite{Magradze1}, cosmic ray physics \cite{Argiro1}, solar physics \cite{Cranmer1}, among others.\\

The Lambert $W$ function\footnote{A generalized Lambert $W$ function was defined by Scott and his collaborators \cite{Scott4,Mezo1}, however, for the time being we shall stick to the original one parameter definition.} is defined mathematically as the multivalued inverse of the function $x\text{e}^{x}$, i.e.,
\begin{equation} \label{lwf}
W(y)\text{exp}[W(y)]=y.
\end{equation} 
If $-\frac{1}{\text{e}}<y<0$, there are two real solutions, and thus two real branches of $W$ \cite{Veberic1}. If complex values of $W$ are allowed, we get many solutions, and $W$ has infinitely many complex branches \cite{Corless1,Corless2,Mezo1}. The earliest mention of Eq. (\ref{lwf}) is attributed to Euler \cite{Euler1}, nevertheless, Euler himself credited Lambert for his earlier work on Lambert's transcendental equation \cite{Lambert1} which has the form
\begin{equation} \label{LTE}
x^m-x^n=(m-n)\nu x^{m+n},
\end{equation}
where $m,n,\nu$ are constants. In fact, Lambert originally developed a series solution (finding $x$ in powers of $q$) of the trinomial equation \cite{Corless1}
\begin{equation} \label{trineq}
x=q+x^{\alpha}.
\tag{2'}
\end{equation}
He later extended the series to give powers of $x$ as well \cite{Lambert1,Lambert2}. Euler \cite{Euler1} transformed Eq. (\ref{trineq}) into the more symmetrical form given in Eq. (\ref{LTE}) by substituting $x^{-n}$ for $x$ and setting $\alpha = mn$ and $q=(m-n)\nu$.\\

The remarkability of the Lambert $W$ function lies in the fact that $W$ is the root of the simplest exponential polynomial function $x\text{e}^{x}=y$. $W(y)$ assumes real values for $y \geq -\frac{1}{\text{e}}$. Three particularly important values of $W(y)$ at $y=-\text{e}^{-1},0,1$ can be computed as $-1,0,0.567143$ respectively. A special name for the last one is the omega constant and can be considered a sort of "golden ratio" of exponentials \cite{http2}. The $n^{\text{th}}$ derivatives of the Lambert $W$ function are given by 
\begin{equation}
W^{n}(y)=\frac{W^{n-1}(y)}{y^n[1+W(y)]^{2n-1}}\varphi _{k=1}^{n} \delta _{kn}W^{k}(y),~~~~y \neq -\frac{1}{\text{e}},
\end{equation}
where $\delta _{kn}$ is the number triangle
\[ \begin{array}{ccccc}
1 & & & &\\
-2 & -1 & & & \\
9 & 8 & 2 & & \\
-64 & -79 & -36 & -6 & \\
625 & 974 & 622 & 192 & 24  
\end{array} .\]
Thus, the first order derivative of $W(y)$ has the expression
\begin{eqnarray}
W'(y) &=& \frac{W(y)}{y[1+W(y)]} \nonumber \\
&=& \frac{\text{e}^{-W(y)}}{1+W(y)}.
\end{eqnarray}
The antiderivative of $W(y)$ can be obtained as
\begin{equation}
\int W(y)\text{d}y=y\left[W(y)-1+\frac{1}{W(y)}\right]+C,
\end{equation}
where $C$ is the arbitrary constant of integration. These were some of the basic properties of the Lambert $W$ function. Additional features of this special function can be found in \cite{Cranmer1,Veberic1,Mezo1,http2}. In spite of such a wide range of applications in different branches of mathematics and physics, its implications in studying the cosmic history of the Universe have never been explored. However, the following two points have motivated us to study this special function in the context of Cosmology---\\\\
$\bullet$ It is well known that the Lambert $W$ function appears in mathematics when one has to solve equations involving a variable which appears both inside and outside of either an exponential function or a logarithm, such as in the equations $\mathrm{e}^x=4x+5$ and $\mathrm{ln}(3x)=x$. Since our Universe has gone through an exponential (inflationary) phase in the past and is presently undergoing a phase of acceleration, similar to the inflationary phase, one cannot help but wonder whether the Lambert $W$ function has some role in the evolution of the Universe. \\\\
$\bullet$ The Lambert $W$ function has appeared implicitly while deriving solutions of the continuity equation in the gravitationally induced adiabatic particle creation scenario \cite{Chakraborty1}. This observation has also motivated us to some extent in studying the importance of this special function in the cosmological perspective.\\

\begin{figure}[t]
\begin{center}
\begin{minipage}{0.4\textwidth}
\includegraphics[width=1.0\linewidth]{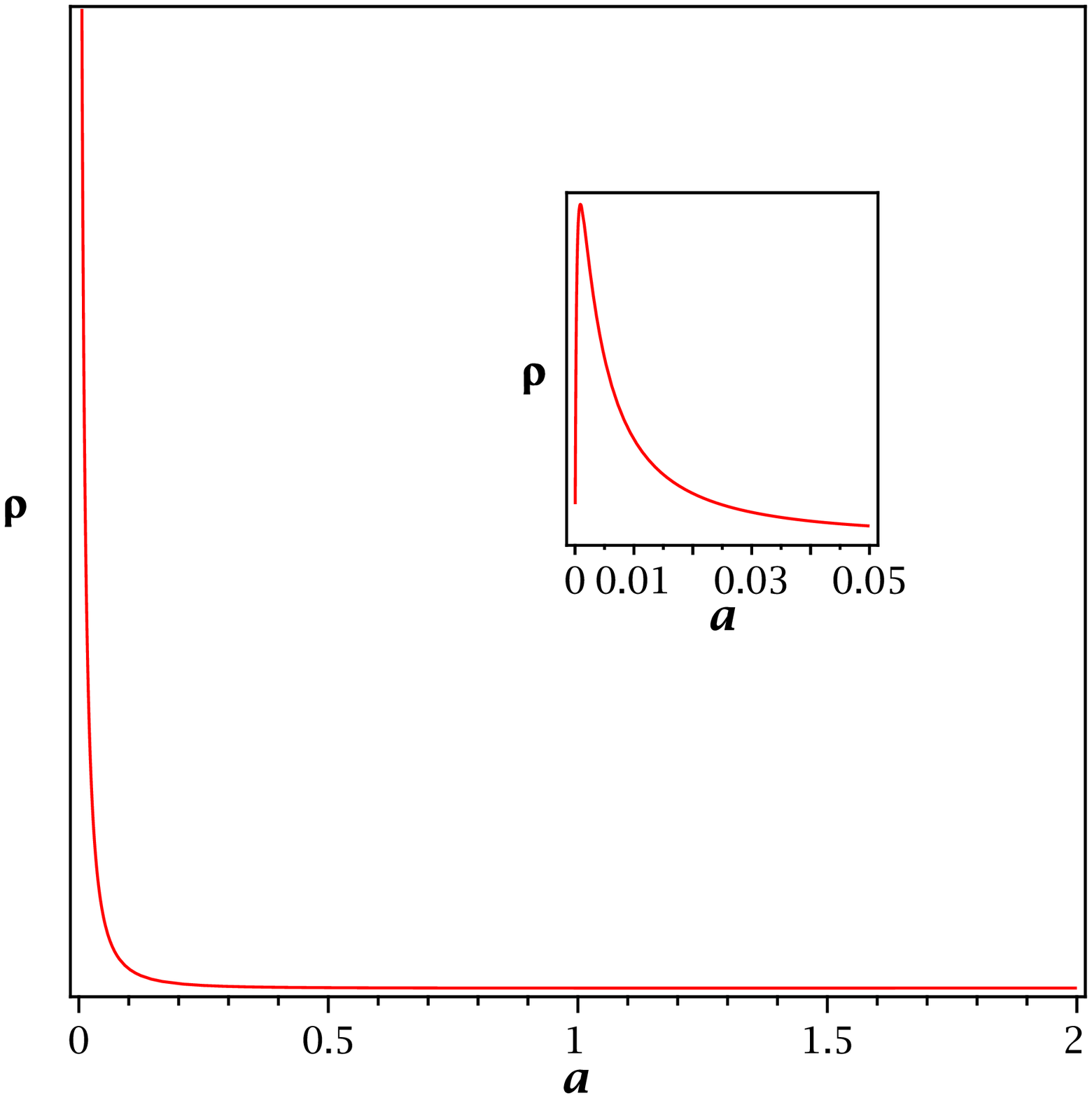}
\end{minipage}
\hspace*{0.5cm}
\begin{minipage}{0.4\textwidth}
\includegraphics[width=1.0\linewidth]{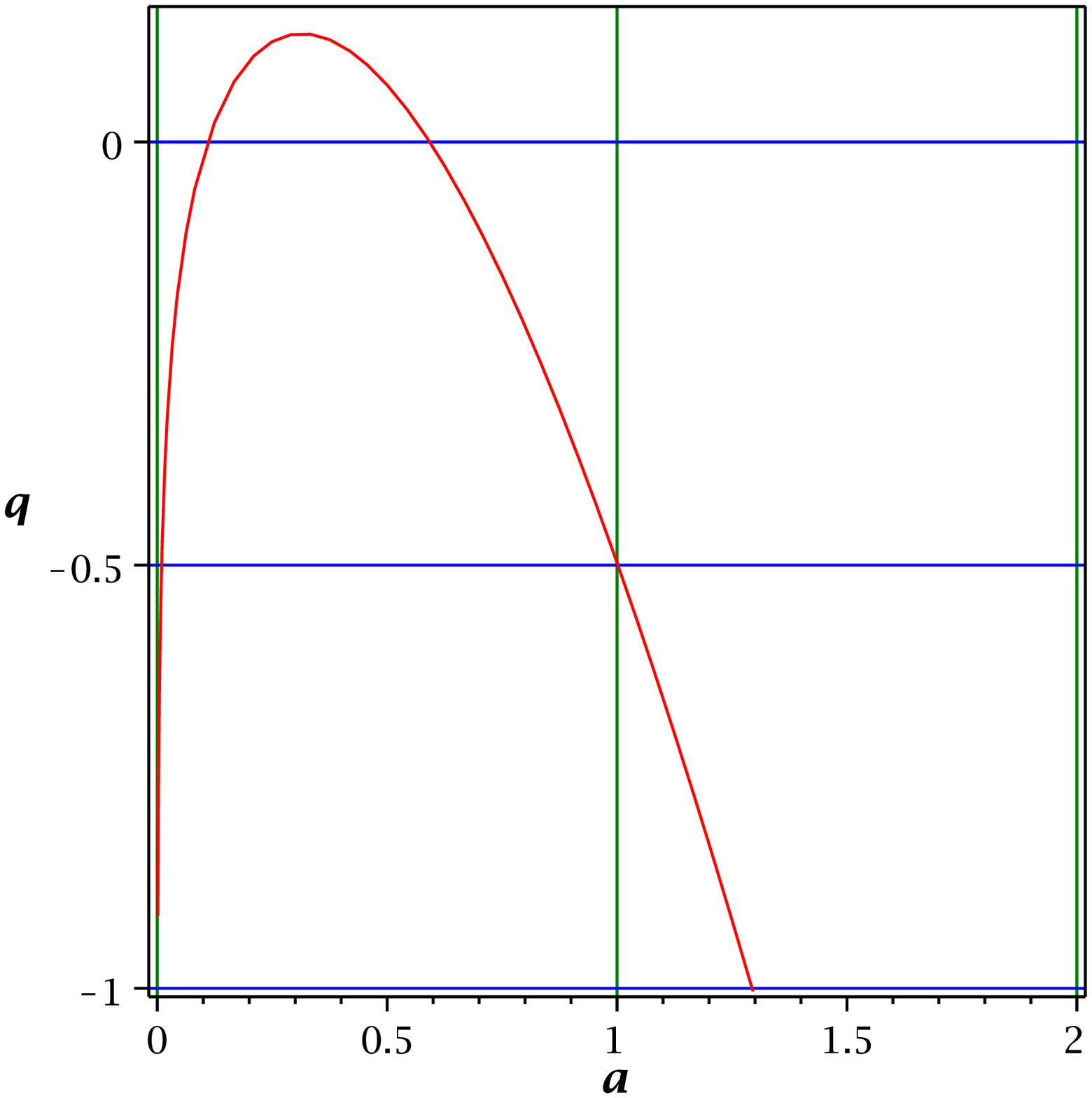}
\end{minipage}
\caption{The left panel shows the variation of the energy density $\rho$ (or equivalently, $H^2$) against the scale factor $a$ along with a magnification in the interval [0,0.05] displayed in inset. The right panel shows the evolution of the deceleration parameter $q$ against $a$. We have considered $\vartheta _\mathrm{1}=\frac{1}{7}$ and $\vartheta _\mathrm{2}=-\frac{16}{5}$.}
\label{Fig:1}
\end{center}
\end{figure}

To start with, let us consider a flat, homogeneous and isotropic Friedmann-Lemaitre-Robertson-Walker (FLRW) universe in comoving coordinates ($t$, $r$, $\varphi$, $\phi$) governed by the metric (assuming $c=1$)
\begin{equation} \label{flrw}
ds^2=-dt^2+a^2(t)\left[dr^2+r^2(d\varphi ^2+\text{sin}^2\varphi d\phi ^2)\right]
\end{equation}
with the associated Friedmann and acceleration equations given by
\begin{equation} \label{fa}
3H^2=8\pi G \rho~~~~~~~~\text{and}~~~~~~~~\dot{H}+H^2=-\frac{4\pi G}{3}(\rho +3P).
\end{equation}
The above equations are obtained by solving the Einstein's field equations 
\begin{equation}
G_{\mu \nu} \equiv R_{\mu \nu}-\frac{1}{2}Rg_{\mu \nu}=8\pi G T_{\mu \nu},
\end{equation}
where the energy-momentum (EM) tensor $T_{\mu \nu}$ is assumed to be given by (due to Weyl's postulate)
\begin{equation}
T_{\mu \nu}=(\rho +p)u_\mu u_\nu + p g_{\mu \nu},~~~~u_{\mu}u^{\mu}=-1.
\end{equation}
Note that $G_{\mu \nu}$ is the well-known Einstein tensor. Now, the energy-momentum (EM) conservation equation is obtained from the contracted Bianchi identity $\nabla_{\mu}T^{\mu\nu}$:
\begin{equation} \label{emce}
\dot{\rho}+3H(\rho +P)=0.
\end{equation}
In Eqs. (\ref{flrw}), (\ref{fa}), and (\ref{emce}), $a(t)$ is the scale factor of the Universe, $H=\frac{\dot{a}}{a}$ is the Hubble parameter, while, $\rho$ and $P$ are, respectively, the energy density and pressure of the cosmic fluid. In order to solve the above equations, we need an EoS connecting $\rho$ and $P$. Now, keeping in line with our previous discussion, we suppose that the EoS of the cosmic fluid is given by
\begin{equation} \label{eos}
w_{\mathrm{eff}} = \frac{P}{\rho} = \left[\vartheta _\mathrm{1}\text{ln}\left\{W\left(\frac{a}{a_0}\right)\right\}+\vartheta _\mathrm{2}\left\{W\left(\frac{a}{a_0}\right)\right\}^3\right],
\end{equation}
where $a_0$ is some positive constant, regarded as the value of the scale factor at the present epoch, while $\vartheta _\mathrm{1}$ and $\vartheta _\mathrm{2}$ are dimensionless parameters which should be fixed from observations. At first sight, the proposed EoS seems to be phenomenological and a bit speculative, but, it is remarkable to know that such a complex EoS can smoothly reproduce all the well-known evolutionary stages of the Universe. Let us now focus on understanding the behaviors of important cosmological parameters such as the Hubble paramter $H$ and the deceleration parameter $q$ due to the consideration of an EoS of the above type. First of all, we set the values of $8\pi G$ and $a_0$ to unity, without any loss of generality. Then, plugging Eq. (\ref{eos}) into the conservation equation (\ref{emce}) and integrating, we obtain the energy density $\rho$ in the following form ($\rho _0$ is a positive constant):
\begin{equation}
\rho = \rho _{0} \text{exp}\left[-3\left\{\text{ln}[W(a)][\vartheta _\mathrm{1} W(a)+\vartheta _\mathrm{1} +1]+W(a)(1-\vartheta _\mathrm{1})+\frac{\vartheta _\mathrm{2}}{12}W(a)^3[4+3W(a)]\right\}\right].
\end{equation}
Furthermore, the deceleration parameter $q$ for the present model is evaluated as
\begin{eqnarray}
q &=& -\frac{\dot{H}}{H^2}-1 \nonumber \\
&=& \frac{3}{2}\left\{1+\vartheta _\mathrm{1}\text{ln}[W(a)]+\vartheta _\mathrm{2}W(a)^3\right\}-1 \label{q-1}\\
&=& \frac{3}{2}\left\{1+\vartheta _\mathrm{1}\text{ln}\left[W\left(\frac{1}{1+z}\right)\right]+\vartheta _\mathrm{2}W\left(\frac{1}{1+z}\right)^3\right\}-1. \label{q-2}
\end{eqnarray}
The last equality expresses the deceleration parameter in terms of the redshift parameter $z$, which connects the scale factor with the relation $a=\frac{1}{1+z}$. Now, on choosing the values of the unknown paramters to be $\vartheta _\mathrm{1}=\frac{1}{7}$ and $\vartheta _\mathrm{2}=-\frac{16}{5}$, our proposed EoS provides us with some interesting consequences in the cosmological perspective. Note that due to a high degree of complexity in the expression for $\rho$, we are unable to solve it for the scale factor $a(t)$. Therefore, in order to realize the behavior of energy density and deceleration parameter at different epochs of evolution, we have plotted\footnote{All the plots in this manuscipt have been produced using Maple software.} the variations of $\rho$ and $q$ against the scale factor $a$. These variations are presented in Figure \ref{Fig:1}. The left panel shows that $\rho \rightarrow +\infty$ as $a \rightarrow 0$, which refers to the initial singularity of the Universe, commonly known as the Big Bang. From the right panel, it is clear that $q$ shows two transitions, both occurring at past redshifts. In other words, the Universe undergoes smooth evolution which starts from an early acceleration phase (inflation), then passes through a medieval deceleration phase (radiation and matter dominated phases), and finally enters into a late acceleration phase. It also shows that the late-time acceleration continues forever. It is worth noting that these deductions are fully consistent with the established evolutionary history of the Universe as demonstrated by the concordance $\Lambda$CDM model of the Universe. A few comments on the choice of $\vartheta _\mathrm{1}$ and $\vartheta _\mathrm{2}$ are in order. At a first look, the values $\vartheta _\mathrm{1}=\frac{1}{7}$ and $\vartheta _\mathrm{2}=-\frac{16}{5}$ may seem to be chosen completely arbitrarily, however, the following two scenarios will establish a motivation for the above choices of the two arbitrary parameters---\\\\
$\bullet$ {\bf Scenario 1:} At the present epoch, we have $a=1$. Plugging this value of $a$ into Eq. (\ref{q-1}) and noting that $W(1)=0.567143$, we obtain the linear equation $-0.8507157038~\vartheta _\mathrm{1}+0.2736333249~\vartheta _\mathrm{2}=-1$. Here, we have used the fact that at the present epoch, $q=-0.5$ as indicated by theoretical prediction of the widely accepted $\Lambda$CDM model and subsequently verified by observations.\\\\
$\bullet$ {\bf Scenario 2:} It is known from observations that the Universe entered the present epoch of cosmic acceleration from the deceleration era at a redshift $z_{da} \approx 0.72$ \cite{Farooq1}. Plugging this value of $z$ into Eq. (\ref{q-2}) and noting that $q=0$ at the transition redshift $z_{da}$, we arrive at a second linear relation $-1.402403228~\vartheta _\mathrm{1}+0.09077772462~\vartheta _\mathrm{2}=-0.5$.\\\\
Solving the above two linear equations, we arrive at the following values of the parameters: $\vartheta _\mathrm{1}=0.1501996764 \approx \frac{1}{7}$, $\vartheta _\mathrm{2}=-3.187560495 \approx -\frac{16}{5}$. Thus, the prior choices of the parameters $\vartheta _\mathrm{1}$ and $\vartheta _\mathrm{2}$ are quite justified.\\

Let us now perform a consistency check on $w_{eff}$ in support of our choice of values for the free parameters $\vartheta _\mathrm{1}$ and $\vartheta _\mathrm{2}$. First of all, note that in the standard, $\Lambda$CDM model of Cosmology, $w_{eff}=\frac{p}{\rho}=\frac{p_m+p_d}{\rho_m+\rho_d}=w_d\Omega_d$, where $p_m$ and $\rho_m$ are the pressure and the energy density of dust, and, $p_d$ and $\rho_d$ are the pressure and energy density of dark energy, $\Lambda$. The last equality is obtained by considering the fractional energy densities of dust and $\Lambda$, given by $\Omega_m=\frac{\rho_m}{\rho}$ and $\Omega_d=\frac{\rho_d}{\rho}$ respectively and noting that $w_d=-1=\frac{p_d}{\rho_d}$ is the EoS of dark energy, $\Lambda$. Now, analysis of recent observations \cite{Aghanim1} suggest that $\Omega_d=0.685$ (since $\Omega_m=0.315$ and $\Omega_m+\Omega_d=1$). Thus, $w_{eff}$ turns out to be $w_{eff}=w_d\Omega_d=-0.685$ at the present epoch. On the other hand, plugging in the values of $\vartheta _\mathrm{1} \approx \frac{1}{7}$ and $\vartheta _\mathrm{2} \approx -\frac{16}{5}$ into Eq. (\ref{eos}) gives $w_{eff} \approx -0.667$ at the present epoch. As one can see, this value is in excellent agreement with the value of $w_{eff}$ obtained from observational data.\\

In summary, in this short paper, we have proposed a novel EoS for the fluid content of a flat FLRW universe which incorporates the Lambert $W$ function in a special fashion [Eq. (\ref{eos})]. Two free parameters, namely, $\vartheta _\mathrm{1} \approx \frac{1}{7}$ and $\vartheta _\mathrm{2} \approx -\frac{16}{5}$ have been introduced and their values have been fixed from the analysis of recent observational data. We have obtained expressions for the energy density $\rho$ and the deceleration parameter $q$ by using the EM conservation equation (\ref{emce}) and the Einstein's field equations (\ref{fa}) respectively. Further, we have plotted the variations of $\rho$ and $q$ against the scale factor $a$ in Fig. \ref{Fig:1}. It is observed that the new EoS proposed by us, although phenomenological and a bit speculative, is successfully able to explain the evolutionary stages of the Universe starting from an early acceleration phase and passing through a deceleration phase before entering into a late-time acceleration phase. The model also depicts that the initial singularity is unavoidable and asserts that the late-time acceleration would continue forever. We have also performed a consistency check on the effective EoS $w_{eff}$ and found that our calculated value is in excellent agreement with that obtained by the analysis of observational data. Therefore, in view of the above discussion, we reiterate that this work represents a small, yet significant step towards employing and understanding the implications of special mathematical functions in the evolution of the Universe. In a future work, we plan to undertake a perturbative analysis and a detailed phase space analysis in order to have a deeper understanding of the proposed model.

\begin{acknowledgments}

The work of KB was supported in part by the JSPS KAKENHI Grant Number JP 25800136 and Competitive Research Funds for Fukushima University Faculty (18RI009 and 19RI017). The authors are thankful to the anonymous reviewer for constructive comments and criticisms which have helped to improve the quality of the manuscript significantly. Author SS would like to dedicate this paper to Professor Subenoy Chakraborty on the auspicious occasion of his $60^{\text{th}}$ birthday.

\end{acknowledgments}


\end{document}